\documentclass[%
 reprint,
 amsmath,amssymb,
 aps,
]{revtex4-2}

\usepackage[T1]{fontenc}
\usepackage{graphicx}
\usepackage{xcolor}
\usepackage{soul}
\usepackage{dcolumn}
\usepackage{bm}
\newtheorem{theorem}{Theorem}

\newcommand{\EqRef}[1]{Eq.~\eqref{#1}}
\newcommand{\FigRef}[1]{Fig.~\ref{#1}}

\newcommand{\MeanT}{\langle T \rangle}

\begin{document}


\title{Escape time in bistable neuronal populations driven by colored synaptic noise}

\author{Gianni Valerio Vinci}
 \email{gianni.vinci@iss.it}
\author{Maurizio Mattia}
\affiliation{Natl. Center for Radiation Protection and Computational Physics \\ 
Istituto Superiore di Sanità, 00161 Roma, Italy}

\date{\today}
\begin{abstract}
Local networks of neurons are nonlinear systems driven by synaptic currents elicited by its own spiking activity and the input received from other brain areas. 
Synaptic currents are well approximated by correlated Gaussian noise. 
Besides, the population dynamics of neuronal networks is often found to be multistable, allowing the noise source to induce state transitions.
State changes in neuronal systems underlies the way information is encoded and transformed. 
The characterization of the escape time from metastable states is then a cornerstone to understand how information is processed in the brain.
The effects of correlated input forcing bistable systems have been studied for over half a century, nonetheless most results are perturbative or valid only when a separation of time scales is present. 
Here, we present a novel and exact result holding when the correlation time of the noise source is identical to that of the neural population, hence solving in a very general setting the mean escape time problem.
\end{abstract}

\maketitle

Recurrent neural networks (RNN) are powerful and widespread models used to study the emergent dynamics of large networks of interacting neurons and their information processing capabilities \cite{Sussillo2014,Vyas2020}. 
The single node dynamics is derived either phenomenologically \cite{Wilson1972,Ostojic2011} or by first-principle under mean-field approximation \cite{Treves1993,Brunel1999,Schaffer2013,Montbrio2015,Mattia2016}.
They generally consist of differential equations for macroscopic quantities that describe the collective behaviour of neuronal populations. 
A common choice is the population firing rates $R(t)$, i.e. the fraction of neurons that emit spikes per unit of time. 
The simplest dynamical law for $R(t)$ is:
\begin{equation}
    \tau_{R} \dot{R}=\Phi[s \,R +H(t)] -R
\label{eq:RNN}
\end{equation}
where $\Phi$ is the current-to-rate gain function, $s$ controls self-excitation, and $H(t)$ is the synaptic current afferent to the neural population. 
$\tau_R$ is the relaxation time that we set to be the unit measure of time ($\tau_R = 1$ in the following).

By bridging micro- and and mesoscopic scales, networks of units like \eqref{eq:RNN} can effectively model brain networks provided that suited connectivity matrices are incorporated.
Besides, RNNs are universal approximators of dynamical systems \cite{Funahashi1993}, and as such they can mimic any computer algorithm \cite{Siegelmann1992,Koiran1994}.
Given such computational power, it is not surprising with that over time RNNs have been widely used as ``reservoir computers'' modeling brain functions like storing memories \cite{Cohen1983,Hopfield1984} and performing cognitive tasks \cite{Mastrogiuseppe2017,Yang2019,Vyas2020}.
Despite the multidisciplinary relevance of this modeling framework, the mechanistic roots of these computational capabilities and their intricate dependency on both network connectivity and single nodes features, is not yet fully understood.
Of particular interest is the case of self-coupling $s>s^*$ for which the neuronal population displays bistability (for $\Phi(x)=\tanh(x)$ the critical value is $s^*=1$). 
Bistability in local cortical network has been linked with neural information processing and has been found in neuronal data as well \cite{Brinkman2022}.
This question prompted us to better understand metastability in these system, in particular the average residence time which will be the main focus of this letter.
In fact, first passage times, i.e. the time needed to jump between states due to fluctuations, are of primary importance in statistical physics \cite{Bray2013,Hanggi1990}, biology \cite{Codling2008}, chemistry \cite{VanKampen1992} and many more fields of science.
The synaptic current $H(t)$ is generally a sum of external stimuli and neuronal activities of other population in the brain. 
Due to various sources of intrinsic randomness \cite{Faisal2008} we can assume $H(t)$ to be a random process \cite{Tuckwell1989}. 
One of the main source of such noise is the fact that neuronal populations have a finite size (i.e. a finite number of neurons) \cite{Brunel1999,Mattia2002}. 
Finite-size fluctuations are correlated in time \cite{Mattia2002,Vinci2023}.
Besides, such fluctuating activity is non-instantaneously delivered to postsynaptic neurons with a distribution of transmission delays further filtered by a non-instantaneous synaptic transmission.
Both these effects can be accurately modeled considering the synaptic current $H(t)$ as an Ornstein-Uhlenbeck (OU) process with correlation time $\tau$ \cite{Mattia2019}.
In summary we will study the stochastic system: 
\begin{equation}
   \begin{split}
   dR & = \left[\Phi(s R + H) -R\right] dt \\ 
   dH &= -\frac{H}{\tau} dt +\frac{\sigma}{\sqrt{\tau}} dW
   \end{split}
\label{eq:NuISis}
\end{equation}
where $dW(t)$ is a standard Gaussian noise ($\langle dW \rangle = 0$, and $\langle dW(t) dW(t') \rangle = \delta(t-t')dt $).
The synaptic input $H(t)$ is an OU process with correlation time $\tau$, zero mean and asymptotic variance $\sigma^2/2$.
\cite{Mattia2019,Mattia2021}.
As mentioned above, for strong enough self-coupling (i.e, $s>1$ for the example case $\Phi(x) = \tanh(x)$), \eqref{eq:NuISis} exhibits bistability.
Numerous results have been found in the years that allowed to better understand how noise-induced transition are affected by the level of correlation in the noise source \cite{Hanggi1995}. 
However, most of them rely on perturbative approaches for $\tau \ll 1$ or $\tau \gg 1$. 
To the best of our knowledge, no exact results have yet been found in the challenging regime where the two time scales are identical, i.e. $\tau=\tau_{R} =1$. 
Here we show that an exact result exists allowing us to work out the statistical features of the residence times $T$ in one of the metastable states, and to discuss how to leverage the newly found result to go beyond the case $\tau=1$ envisaging possible applications.

\paragraph*{The Fokker-Planck operator $\mathcal{L}$ ---}  Our starting point is the Fokker-Planck equation associated to \eqref{eq:NuISis}: 
\begin{equation}\label{eq:FPoriginal}
   \partial_t P = - \partial_{r}[(\Phi(s \, r +h)-r)P] + \frac{1}{\tau} \partial_{h}(h \, P) +\frac{D}{\tau} \partial_h^2P 
\end{equation}
with diffusion coefficient $D = \sigma^2/2$. 
Since we cannot write the drift as a gradient of a potential $\nabla U(r,h)$, \EqRef{eq:FPoriginal} is in general difficult to solve.
To make some progress we perform the following change of variables: $x=s r + h$ and $y = h$. 
Using Ito's lemma $dx=s \, dr +dh$ the following stochastic differential equations (SDE) result:
\begin{equation}
    \begin{split}
    dX &= \left[s\,\Phi(X) - X + \left(1 - \frac{1}{\tau}\right) Y\right]dt +\frac{\sigma}{\sqrt{\tau}} dW\\
    dY &=-\frac{Y}{\tau} dt +\frac{\sigma}{\sqrt{\tau}} dW
    \end{split}  \, .
\label{eq:XYSis}
\end{equation}
For the specific case $\tau=\tau_\mathrm{R} = 1$, such stochastic dynamics simplifies to
\begin{equation}
   \begin{split}
      dX & = \left[s\Phi(X)-X\right] dt + \sigma \, dW\\
      dY & = -Y \, dt + \sigma \, dW
   \end{split} \, ,
\label{eq:XYSisFinal}
\end{equation}
where the two variables $x$ and $y$ are not independent only because they are driven by the same white noise $\sigma W(t)$.
The associated Fokker-Planck equation is:
\begin{equation}\label{eq:FPx}
\begin{split}
 \partial_t P  &= \mathcal{L} P \\
 &= (\mathcal{L}_x +\mathcal{L}_y +\mathcal{L}_{xy})P
\end{split}
\end{equation}
with $\mathcal{L}_x \cdot= -\partial_x \left[(s\Phi(x)-x) \cdot \right] + D \, \partial_x^2 \cdot$, $\mathcal{L}_y \cdot = \partial_y(y \cdot) + D \, \partial_y^2 \cdot$  and $\mathcal{L}_{xy} \cdot = 2D \, \partial_x\partial_y\cdot$.
Here we remark that the operator $\mathcal{L}_x$ is the generator of the one-dimensional stochastic process
\begin{equation}
    dX=\left[ s\Phi(X)-X \right] dt +\sigma \, dW \, ,
\label{eq:XSis}
\end{equation}
which in what follows we will take as reference process in characterizing the residence times of the original two-dimensional system \eqref{eq:NuISis}.
Despite the fact that in absence of noise ($\sigma = 0$) the drift of (\ref{eq:FPx}) is the gradient of a potential function of two independent variable, detailed balance is  still broken by the mixed derivatives in $\mathcal{L}_{xy}$ \cite{Deniz2016}.
However, we have now access to the marginal distributions of both variables under stationary conditions (i.e., $\partial_t P = 0$) defined as $f_0(x) = \int_{\mathcal{D}_y} P(x,y)dy$ and $g_0(y) = \int_{\mathcal{D}_x} P(x,y)dx$.
Indeed, due to the natural boundary conditions of the system (i.e., vanishing probability currents and densities for $|x|, |y| \to \infty$), by integrating \EqRef{eq:FPx} with $\int_{\mathcal{D}_y} dy$, we obtain $\mathcal{L}_x f_0(x)=0$ whose solution is the Gibbs measure:
\begin{equation}
    f_0(x)=\frac{e^{-U(x)/D }}{\int_{\mathcal{D}_x} e^{-U(x)/D}dx} \, .
\label{eq:f0}
\end{equation}
where $U(x) = \int \left[s \Phi(x)-x \right] dx$ is the potential for $x$ in \EqRef{eq:XSis}.

\paragraph*{Dominant eigenvalues of $\mathcal{L}$ ---} 
Since the introduced transformation of variables in the previous section is linear, the eigenvalues of the original operator with variable $r$ and $h$, are the same as the operator $\mathcal{L}$ in the variable $x$ and $y$.

The decay rate is the sum of escape rates from all the minima of the energy landscape \cite{Risken1996}.
In what follow we will assume, without loss of generality, that $U(x)$ is a bistable symmetric potential hence we have that the  mean escape time  is $\MeanT = 2/|\lambda_1|$ were $\lambda_1$ is the dominant (closest to zero) eigenvalue of $\mathcal{L}$.
We can now prove, for arbitrary $\Phi(x),s$ and $\sigma$ the following :
\begin{theorem}
     If $\mathcal{L}=\mathcal{L}_x +\mathcal{L}_y +\mathcal{L}_{xy} $ with $\mathcal{L}_{xy}=k\partial_x \partial_y$ and $k$ a real  number, then the spectra of $\mathcal{L}_x$ and $\mathcal{L}_y$ belong also to the spectrum of $\mathcal{L}$ .
\end{theorem}
  
To prove the theorem we will use a pseudo-spectral approach \cite{shizgal2015spectral}, which consists of an expansion of the eigenfunctions of $\mathcal{L}$ in terms of the pairwise product of the eigenfunctions of $\mathcal{L}_x$ and $\mathcal{L}_y$. 
First, we define:

\begin{equation}
\begin{split}
    \mathcal{L}_x f_n & = \lambda_n^{(x)} f_n \\
    \mathcal{L}_y g_n & = \lambda_n^{(y)} g_n
\end{split}
\label{eq:SpectralEqs}
\end{equation}

and we indicate with a tilde $\tilde{}$ the eigenfunctions of the adjoint operators $\mathcal{L}_{x}^{\dagger}$and $\mathcal{L}_{y}^{\dagger}$.
The eigenbasis in braket notation is then: 
\begin{equation}
    \begin{split}
          \langle i,j | &=\tilde{f}_i(x) \tilde{g}_j(y)\\
          |i,j\rangle &=f_i(x) g_j(y)
    \end{split}
\label{eq:xyeigenfunc}
\end{equation}
which is orthogonal with respect to the inner product $\langle \dots \rangle =\int_{- \infty}^{\infty} \dots dxdy$ properly normalized such that $\langle k,m |i,j\rangle = \delta_{k,i}\delta_{m,j}$.  Then the operator $\mathcal{L}$ is isospectral to the infinite matrix:
\begin{equation}\label{eq:MatrixL}
   L_{nm} =\langle n |\mathcal{L} | m\rangle
\end{equation}
where the integer indexes $n$ and $m$ are mapped for convenience to the unique pairs $(i_n,j_n)$ and $(i_m,j_m)$, respectively. 
Direct substitution of $\mathcal{L}$ in \EqRef{eq:MatrixL} leads to:
\begin{displaymath}
   L_{nm}=\Lambda_{nm} + D K_{nm} \, ,
\end{displaymath}
where $\Lambda_{nm} = \delta_{i_m,i_n}\delta_{j_m,j_n}(\lambda_{i_m}^{(x)} + \lambda_{j_m}^{(y)})$ while $D K_{nm} = \langle n |\mathcal{L}_{xy} | m\rangle$.
Consider now the rows of $K_{nm}$ associated to the states $\langle \bar{n}|$  equal to either $\langle i,0 |$ or $\langle 0,i |$ for any arbitrary $i$. 
Then, since both $\tilde{g}_0(y)=\tilde{f}_0(x)=1$:
\begin{equation*}
\begin{split}
K_{\bar{n}m} &=2 \langle \bar{n} |\partial_x \partial_y | m\rangle \\
&=2 \int \partial_x \tilde{f}_{i_{\bar{n}}}(x) f_{j_m}(x) dx \int \partial_y \tilde{ g}_{i_{\bar{n}}}(y) g_{j_m}(y)dy=0  
\end{split}
\end{equation*}
for any arbitrary $m$. 
Hence the rows $\bar{n}$ of $L$ have only one non-zero element $L_{\bar{n}m}=\Lambda_{\bar{n}m}$ which implies that the spectra of $\mathcal{L}_x$ and $\mathcal{L}_y$ are contained in the spectrum of $\mathcal{L}$.
Indeed, versors $\vec{e}_{\bar{n}}$ with elements $\delta_{m\bar{n}}$ are eigenvectors of $\mathbf{L}$ as $\mathbf{L}\vec{e}_{\bar{n}} = \Lambda_{\bar{n}\bar{n}} \vec{e}_{\bar{n}}$.
This result is true for any arbitrary well-defined operators  $\mathcal{L}_x$ and $\mathcal{L}_y$ $\blacksquare$.

\begin{figure*}
    \centering
    \includegraphics{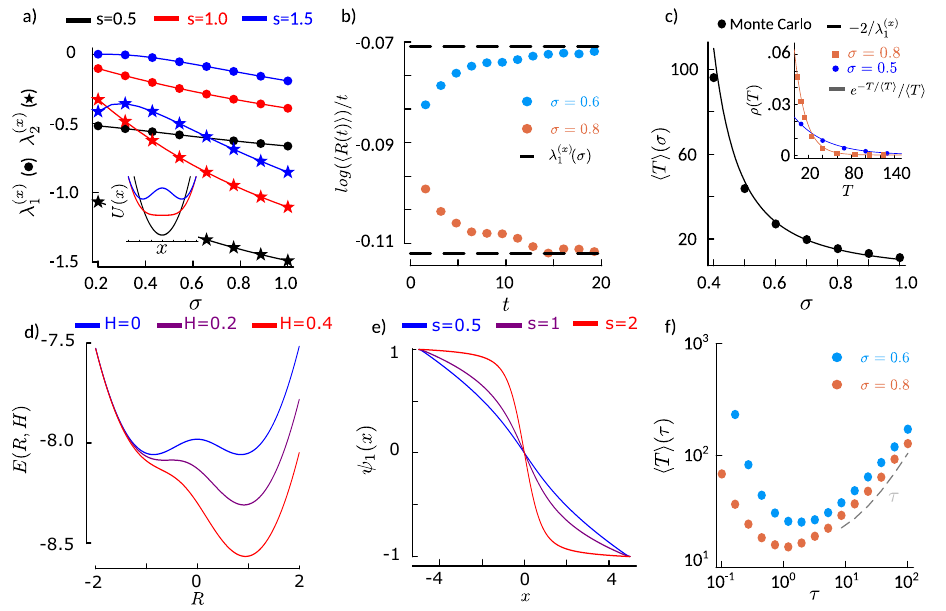}
    \caption{a) Dominant ($\lambda_1^{(x)}$, circle) and sub-dominant ($\lambda_2^{(x)}$, star) eigenvalues of $\mathcal{L}_x$ as function of the size $\sigma$ of the synaptic-input fluctuations for $s = \{0.5, 1, 1.5\}$ (black, red and blue, respectively). 
    Eigenvalues are the numerical solutions of the spectral \EqRef{eq:SpectralEqs} for $x$.
    Inset, potential $U(x)= - \int(s \Phi(x) - x) dx$ for the same $s$. 
    In panels (b-d) and (f) $s=1.5$ to have bistability.
    b) Relaxation of the normalized firing rate $R(t)/t$ averaged across $10^5$ independent numerical integrations of \EqRef{eq:NuISis} with $R(0)$ randomly sampled from a normal distribution with zero mean and unitary variance.
    Dashed lines, $\lambda_1^{(x)}$ for the sampled $\sigma =$ 0.6 and 0.8 (azure and brown, respectively).
    c) Average escape time $\MeanT$ of $10^4$ independent Monte Carlo simulations (numerical integrations of \EqRef{eq:NuISis}) and $-2/\lambda_1^{(x)}$.
    $R(0)$ are randomly set in one of the two minima of $U(x)$ and $H(0)=0$.
    Inset, distributions $\rho(T)$ of escape times $T$ estimated from simulations with $\sigma = 0.5$ and 0.8 (blue and brown dots, respectively) matching the exponential distributions (solid) with expected value $-2/\lambda_1^{(x)}$.
    d)The potential function $E(R,H)=\int^{R} (\Phi(sy +H) -y)dy$ of system (\ref{eq:NuISis}) assuming constant current $H$ (colors indicates different values of $H$).
    e)The dominant adjoint eigenfunction $\psi_1(x)$ at fixed $\sigma=0.8$ is bounded for all value of $s$. $\psi_1$ is computed numerically assuming the domain is bounded in $[-L,L]$ with $L=5$.
    f) Mean escape time as a function of $\tau$ and $\sigma$ (different colors) sampled from Monte Carlo simulation of $10^4$ realizations.
    All numerical integration of SDE have been performed with $\tau=1$ and $dt=10^{-5}$ using Euler–Maruyama method.}
\label{fig:EigenvaluesMFPT}
\end{figure*}

In our case we have that $\lambda_{n}^{(y)}=-n$ and $\tilde{g}_n(y)=H_n(y/\sqrt{2D})$ where $H_n(y)$ are the Hermite polynomials \cite{Risken1996}, 
while $\lambda_n^{(x)}$ can be either computed numerically, by solving directly the eigenfunctions boundary value problem with the generalized Scharfetter-Gummel method \cite{Augustin2017} (see \footnote[3]{Julia script for numerical computation of eigenvalues and eigenfunctions can be found in the repository: https://github.com/giav1n/NeuralPopColoredNoise} 
for our implementation in Julia), or using analytical approximations, for the dominant eigenvalue, that involves integrals of $f_0 (x)$ \cite{Risken1996}.
All our conclusion are independent on the specific functional form of the gain function $\Phi(x)$.
As case study in this Letter we use the antisymmetric function $\Phi(x)=\tanh(x)$.

It should be noted that in the spectrum $\mathcal{L}$ are also present eigenvalues different from $\lambda_{n}^{(x)}$ and $\lambda_{n}^{(y)}$. 
However two arguments can be made here.
First note that, thanks to the property of Hermite polynomial the diagonal of $\mathbf{K}$ is null (i.e., $K_{nn}=0$), as the partial derivative $\partial_y g_n \propto g_{n+1}$ being $g_n(y)=e^{-y^2/2D}H_n(y/\sqrt{2D})$.
Applying a well know result from eigenvalue perturbation theory we have for the eigenvalues of $\mathbf{L}$:
\begin{equation*}
\begin{split}
       \lambda_n(\mathbf{L}) &= \lambda_n(\mathbf{\Lambda}) +D\frac{\vec{e}_n^{T}\mathbf{K} \vec{e}_n}{\vec{e}_n^{T}\vec{e}_n} + \mathcal{O}(D^2) \\
       &= \lambda_n(\mathbf{\Lambda})  + \mathcal{O}(D^2)
\end{split} \, ,
\end{equation*}
where the versors $\vec{e}_n$ are the eigenvectors of the diagonal matrix $\mathbf{\Lambda}$.
Consequently, the full spectrum of $\mathcal{L}$ includes also all the possible sums $\lambda_{i}^{(x)} + \lambda_{j}^{(y)}$ plus perturbation of order $D^2$.
As $\lambda_n(\mathbf{\Lambda})$ depend themselves on $D$, here we are not excluding the possibility that such eigenvalues are not the leading terms of the expansion.
Moreover, from a physical point of view, it is typical of multistable systems in regime of small fluctuations to have the first non-null $\mathrm{Re} \, \lambda_1 \ll 1$  which is a clear manifestation of the separation of time scales induced by the potential barrier that separates different stable states.
Since for the one-dimensional $x$-system $\lambda_n^{(x)} \ll 1$, as shown in\FigRef{fig:EigenvaluesMFPT}a, we conjecture that the dominant eigenvalue of $\mathcal{L}$ for the full system \eqref{eq:NuISis}, and thus \eqref{eq:XYSis}, is precisely $\lambda_1 = \lambda_{1}^{(x)}$.
We test this conjecture in numerical simulations.
By setting an initial condition of the system relatively close to a stationary and stable solution, a generic expectation value $m(t) = \langle h(r,h)| P(r,h,t) \rangle$ of an arbitrary function $h(r,h)$ follows the exponential decay
\begin{displaymath}
    m(t)\sim m(\infty) + [m(0)-m(\infty)] e^{\lambda_1 t} \, ,
\end{displaymath} 
where $m(\infty)$ is the expectation value at equilibrium. 
This asymptotic limit allows to estimate the dominant eigenvalue as 
\begin{equation}
    \lambda_1 = \lim_{t \to \infty}\frac{1}{t} \log \frac{m(t)-m(\infty)}{m(0)-m(\infty)} \, .
\label{eq:alternativeLambda}
\end{equation}
In \FigRef{fig:EigenvaluesMFPT}b the approximations to such limit are shown for the average firing rate $m(t) = \langle R(t) \rangle = \langle r | P(r,h,t)\rangle$, which for sufficiently large $t$ remarkably match the $\lambda_1^{(x)}$ of the $x$-system \eqref{eq:XSis} (dashed lines).

In \FigRef{fig:EigenvaluesMFPT}c we show the excellent agreement between the mean escape time $\MeanT$ of Monte Carlo simulations of system (\ref{eq:NuISis}) for $\tau=1$ and $2/|\lambda_1|$.
The equivalence of escape statistics between systems (\ref{eq:XSis}) and (\ref{eq:NuISis}) goes beyond the mean as expected for such a separation of time scales eventually leading to a quasi-exponential distribution $\rho(T)$ (\FigRef{fig:EigenvaluesMFPT}c-inset).

\paragraph*{Beyond $\tau=1$ ---}  
When $\tau \ll 1$, the synaptic input $H(t)$ approaches the white noise limit and the firing rate $R(t)$ in \EqRef{eq:NuISis}) is no longer an Ito equation because the noise $W(t)$ enters nonlinearly in the drift term of $R$.
However, in the limit of fast synaptic input $\tau \to 0$, $\tau dy \to 0$ in \EqRef{eq:XYSis} and $y$ approaches a Wiener process: $y \, dt = \sigma \sqrt{\tau} dW$.
Replacing this expression in the SDE \eqref{eq:XYSis} for $x$, we eventually obtain the one-dimensional stochastic dynamics
\begin{displaymath}
  dX=\left[ s \, \Phi(X) - X \right]dt + \sigma \sqrt{\tau} dW \, .  
\end{displaymath}
For this systems the mean escape time is given by the inverse of the usual Kramers' escape rate \cite{Hanggi1990,Risken1996}: $\langle T \rangle \approx e^{2\Delta/\sigma^2 \tau}$ with $\Delta= U(x_M) -U(x_m)$ and $x_{M}$($x_{m}$) is the position of the maximum (minimum) of the potential $U(x)$.
Given the scaling rule chosen for the synaptic noise ($\sigma \sqrt{\tau}$), in the limit $\tau \to 0$ it vanishes and the likelihood to cross the barrier of $U(x)$ reduces to 0 as $e^{-1/\tau}$.

In the opposite regime, $\tau \gg 1$, we can use the argument presented in \cite{MorenoBote2004,Woillez2020}. 
In fact in this case we have that $R$ is much faster then $H$ and we proceed with an adiabatic approximation. 
In other words we can assume that $R$ reaches its stationary value $R_s$ instantaneously hence, $\Phi(s R_s + H) - R_s = 0$. 
We can rewrite it as the input $H$ needed to reach in a time window smaller than $\tau$, the asymptotic value $R_s$:
\begin{displaymath}
  H = \Phi^{-1}(R_s)-s \, R_s  \, .
\end{displaymath}
As shown in \FigRef{fig:EigenvaluesMFPT}d, for small enough $H$, bistability is preserved and no escapes are allowed.
A transition deterministically occurs only when the input is larger than a threshold value $H_M$, i.e., when only one attractor $R_s$ exists, and it is beyond the barrier of the unperturbed system at $R = 0$.  
Consequently, the escape time in this regime is given by the probability to have $H>H_M$ in time windows $\mathcal{O}(\tau)$.
As fluctuations of $H$ have constant variance $D$, the first-passage time needed for such OU process to reach $H_M$ is asymptotically linear in $\tau$: $\MeanT \approx \tau$ \cite{Sato1977}. 

Close to $\tau=1$ we can rely on perturbation theory and the results of this Letter. 
It is convenient to consider as a small parameter $\epsilon = 1 -1/\tau$.  
For arbitrary $\tau$ we can write the Fokker-Planck operator as {$\mathcal{L} = \mathcal{L}^{(0)} + \epsilon \mathcal{L}^{(1)}$, where $\mathcal{L}^{(0)}$ is the operator at $\tau=1$ and $\mathcal{L}^{(1)}= -(D \partial_{x}^{2} +y \partial_x +2D\partial_x \partial_y +\mathcal{L}_y^{(0)})$.
Following standard perturbation theory of quantum mechanics, we expand eigenfunctions and eigenvalues as $|n\rangle \sim |n^{(0)}\rangle + \epsilon |n^{(1)}\rangle$, $\lambda \sim \lambda^{(0)} +\epsilon \lambda^{(1)}$, eventually obtaining for the first-order term in $\epsilon$, $\lambda^{(1)} = \langle n^{(0)}|\mathcal{L}^{(1)}|n^{(0)}\rangle $.
Considering $|n^{(0)}\rangle = |i_n,j_n\rangle$ from \EqRef{eq:xyeigenfunc}, the only integrals associated to the four terms of $\mathcal{L}^{(1)}$ differing from 0 are $\langle n^{(0)} | \mathcal{L}_y^{(0)} |n^{(0)}\rangle = \langle j_n|\mathcal{L}_y^{(0)}|j_n \rangle =\lambda_n^{(y)}$ and $\langle n^{(0)} | D\partial_x^2 |n^{(0)}\rangle =D \langle i_n|\partial_x^2|i_n\rangle$.
From this, the $\lambda_1$ contributing to the longest timescale, is given by the eigenmodes minimizing these two perturbative terms. 
For the former this happens when $\lambda_n^{(y)}=-j_n=0$, while for the latter when the eigenfunction $\mathcal{L}_x| 1_x \rangle =  \lambda_1^{(x)}| 1_x \rangle$ with the smallest non-zero eigenvalue $\lambda_1^{(x)}$ is taken into account, eventually leading to
\begin{equation}
    \lambda_1 (\tau) \underset{\tau \to 1}{\sim} \lambda_1|_{\tau=0} - (1-1/\tau)  D \langle 1_x |\partial_{x}^2 |1_x \rangle + \mathcal{O}(\epsilon ^2) \, .
\label{eq:PerturbLambda}
\end{equation}
The derivative of the mean escape time at $\tau=1$ is then $\partial_{\tau} \MeanT|_{\tau=1} = -D \langle 1_x |\partial_{x}^2 |1_x \rangle /\lambda_1^2|_{\tau=0}$. 
Moreover we can demonstrate the following:
\begin{theorem}
    For Fokker-Planck operator $\mathcal{L}$, defined in the closed finite interval $[-L,L]$ with continuous drift and diffusion and natural boundary condition in $\pm L$, the integral $\langle 1_x |\partial_{x}^2 |1_x \rangle \geq 0$.
\end{theorem}
The proof follows form the boundness of the adjoint eigenfunctions (\FigRef{fig:EigenvaluesMFPT}e) \footnote{Since the drift and diffusion are continuous, the adjoint eigenfunctions $\psi_n(x)$ are continuous as well in the finite interval $[-L,L]$. The extreme values theorem guarantees the existent of a constant $C$ such that $\psi_1(x) \geq C$. Moreover thanks to the natural boundary condition $\partial_x \phi_1 (x)|_{\pm L}=0$. Thus $\langle 1 |\partial_{x}^2 |1 \rangle \geq C \int_{-L}^{L}\partial^2_x \phi_1(x)dx=0$ }. Consequently, for rather general systems, $\partial_{\tau} \MeanT|_{\tau=1} \leq 0$  and we can conclude that at least a minimum of $\MeanT(\tau)$ must exists and, if we assume that is unique, it must occur for $\tau \geq 1$. Note that the existence of optimal value of memory has been reported recently also in the context of active Brownian particle \cite{Caprini2021,Dabelow2021,Caprini2021b} where the noise source acts on the drift of the other variable linearly contrary to the case of neural population dynamics. Thus we have analyzed all possible regimes of correlation time and our prediction are in accordance with Monte-Carlo simulations, see \FigRef{fig:EigenvaluesMFPT}f).

\paragraph*{Conclusions ---} Most of the result on escape time under the effect of colored noise are derived as effective limit of white noise where the dimensionality of the problem is reduced.  
On the contrary, our result is exact when the time scales of the two process are identical, $\tau=1$, the most unperturbative regime possible. 
Our conclusions hold for arbitrary value of $s$, $\sigma$ and gain function $\Phi$. 
Moreover, systems like \EqRef{eq:NuISis} are rather prototypical dissipative models that have been used in several field of science.
For this reason we anticipate that our results might be beneficial also outside the neuroscience domain. 
Indeed, one possible application of our results is system identification, where the current $H(t)$ could be experimentally manipulated and network properties like the level of excitability can be inferred from the escape rate. 
Another important application can be found in networks of bistable units proven to be successful in performing cognitive tasks \cite{Stern2023}. 
In fact, some of the information processing capabilities of such systems could be deeply linked to the single unit stability to internal noise, which in RNNs with random connectivity naturally arise as chaotic dynamics \cite{Sompolinsky1988,Stern2014}.

Work partially funded by the Italian National
Recovery and Resilience Plan (PNRR), M4C2, funded
by the European Union - NextGenerationEU (Project
IR0000011, CUP B51E22000150006, ‘EBRAINS-Italy’) to M.M.

\bibliography{refs}

\end{document}